\documentstyle[aps,multicol,epsf]{revtex}
\begin{document}
%\draft
%\preprint{}.
\title{Conformational Instability of Rodlike Polyelectrolytes due to Counterion
Fluctuations}
\author{Ramin Golestanian}
\address{Institute for Advanced Studies in Basic Sciences, Zanjan
45195-159, Iran \\
and Institute for Studies in Theoretical Physics and Mathematics,
P.O. Box 19395-5531, Tehran, Iran}
\author{Tanniemola B. Liverpool}
\address{Condensed Matter Theory Group, Blackett Laboratory, Imperial College,
London SW7 2BZ, UK}
\date{\today}
\maketitle
\begin{abstract}
The effective elasticity of highly charged stiff polyelectrolytes
is studied in the presence of counterions, with and without added
salt. The rigid polymer conformations may become unstable due to
an effective attraction induced by counterion density
fluctuations. Instabilities at the longest, or intermediate length
scales may signal collapse to globule, or necklace states,
respectively. In the presence of added-salt, a generalized
electrostatic persistence length is obtained, which has a
nontrivial dependence on the Debye screening length. It is also
found that the onset of conformational instability is a re-entrant
phenomenon as a function of polyelectrolyte length for the
unscreened case, and the Debye length or salt concentration for
the screened case. This may be relevant in understanding the
experimentally observed re-entrant condensation of DNA.

\end{abstract}
\begin{multicols}{2}

\section{Introduction and Summary}  \label{sIntro}

A polyelectrolyte (PE) is an ionic polymer which when dissolved in
polar solvents dissociates into a long polymer chain (macroion), and
small mobile {\it counterions}. Because of the electrostatic repulsion
of the uncompensated charges on the polymer, the chain is stretched
out to rodlike conformations. Naively, one may expect that
higher charges on the polyelectrolyte lead to more rodlike configurations.
This is not the case, as the tendency towards extended shapes is
opposed by stronger attraction to the counterions.
The latter may condense  on a highly charged polyelectrolyte, giving it a much
lower apparent charge\cite{Man,Oos},  and resulting in a lower stiffness.
Nonetheless, in a mean-field (Poisson-Boltzmann) treatment of the counterions,
the conformation of the polyelectrolyte is stretched\cite{BJ1,LeBret}.

On the other hand, it is now well known from experiments
\cite{Delsanti,huber}, simulations \cite{SK}, and analytical
theories \cite{Monica,BKK,SchP}, that highly charged flexible
polyelectrolytes (such as polystyrene sulphonate) can collapse in
the presence of multivalent counterions to highly compact states.
Stiff polyelectrolytes (such as DNA) could also collapse
\cite{Bloomfield}, and rodlike polyelectrolytes may form bundles
\cite{Bundle,lyubart,stevens}, under similar conditions. To
account for these effects, an attractive interaction capable of
competing with the residual coulomb repulsion (of the
polyelectrolyte backbones with compensated charge densities) is
needed. Note that the situation is more subtle for stiff
polyelectrolytes due to their intrinsic rigidity
\cite{Man2,Rouzina,GKL,Podgornik,Andelman}.

To understand the origin of the attractive interaction, consider
two like-charged substrates on which counterions are condensed. As
the two objects approach each other, the counterions may rearrange
their positions. Any {\em correlated separation of charges} now
leads to an {\it attraction} whose range is of the order of the
``correlation hole'' of the counterions on each substrate. Several
different mechanisms may lead to correlated charge separation: (i)
At low temperatures, the counterions form a Wigner crystal on the
charged substrate, which leads to an attractive interaction with a
range set by the lattice spacing
\cite{Rouzina2,Gron,Shklovskii,Levin,Solis}. (ii) Specific binding
of counterions to the substrate is another mechanism for charge
separation \cite{Leikin}. In this case the periodicity, and hence
the range of the attraction, is dictated by the underlying
structure of the substrate, such as the double helical structure
of DNA along which the charged phosphate groups are located. (iii)
Thermal fluctuations can also induce instantaneous charge
separations which inter-correlate on the two objects, leading to
an attraction similar to the van der Waals interaction. Since the
dominant fluctuations have wavelengths of
the order of the separation of the objects, the range of the
attraction is set by the distance between them; i.e. the
interaction is long-ranged \cite{Oos,BJ1,Gron,Pod1,HaLiu,KGRMP}.
(iv) Finally, it has been proposed that macroions can be
overcharged by the condensing counterions due to the gain in
correlation energy \cite{overcharge}. Consequently, when two such
macroions are brought close to each other in a neutralizing
solution, the distribution of the counterions can be asymmetric
leading, at each instant, to one overcharged and one undercharged
macroion \cite{messina,ray}. This mechanism will lead to
instantaneous ordinary long-ranged Coulomb attraction between the
two decorated macroions, and is reminiscent of covalent
bonding in atomic systems in that the macroions are effectively
sharing a fraction of their condensed counterions \cite{ray}.

As thermal fluctuations clearly set the energy scale for the
relevance of these different attractive interactions, one may
naturally ask the following question \cite{Roland,Andy}: Is room
temperature `high' or `low' for  DNA or actin condensation, or
bundle formation? In regards to this question one should note that
although an actual Wigner crystal may appear far--fetched on the
surface of DNA or actin at room temperature, remnants of such a
structure may survive for multivalent counterions.
%, as has been
%recently observed for the case of actin with heavy counterions in
%an X-ray scattering experiment \cite{Gerard}.

Since it is just the local charge separation that is needed for the
`zero temperature' mechanism, and not the long-range order of the
global structure, a liquid phase should still lead to attraction
provided that some short-range order persists. Consequently, it seems that
both types of interactions are in effect at room temperature and the
dominant mechanism should be determined by comparing the range of the
interaction and the distance between macroions. While the short-ranged
interactions seem to play the dominant role in bundle formation
\cite{Rouzina2,Gron,Shklovskii,Levin,Solis,Leikin,HaLiu}, there seems
to be a consensus that they are not capable of actually collapsing a
single DNA or actin chain \cite{Rouzina,Nguyen}, and thus the
long-ranged interaction caused by thermal fluctuations may be
responsible for the observed condensation\cite{Bloomfield,GKL}.

In this paper, we focus on the long wavelength behavior of
polyelectrolytes. Our approach is complementary to that of the
``Wigner liquid'' theories~\cite{Rouzina,Nguyen}, whose local
formulation corresponds to examination of short wavelength
behavior. They find that the softening contribution to the
persistence length due to the Wigner liquid correlations is not
large enough to fully compensate the mechanical stiffness and
render a null persistence length, which would trigger chain
collapse. We also find that the softening contribution to the
rigidity due to charge fluctuations at the smallest length-scales
is never strong enough to completely negate the bare bending
rigidity, and obtain collapse {\em at these scales}, but rather
that the `collapse' occurs at higher length-scales due to {\em
longer} wavelength charge fluctuations. Additional correlations at
short length-scales as might be obtained with a Wigner liquid
should not affect the large-scale behavior which is the domain of
our theory.

To this end, we employ path integral methods\cite{KGRMP,LiK} to
study the energy cost of deforming a stiff and highly charged
polyelectrolyte in the presence of thermally fluctuating
counterions \cite{GKL}. In particular, consider a chain of length
$L$, with a microscopic persistence length $\ell_p$, and average
separation $a$ between charges on its backbone, in a neutralizing
solution of counterions of valence $z$. The polyelectrolyte is
considered highly charged when $a$ is less than the Bjerrum length
$\ell_B=e^2/\epsilon k_B T$, where $\epsilon$ is the dielectric
constant of the solvent. (For water at room temperature,
$\ell_B\simeq 7.1 \AA$.) We calculate the effective free energy of
a fluctuating polyelectrolyte as a perturbative expansion in the
deformations around an average rodlike structure. The linear
stability of this structure is controlled by a spectrum ${\cal
E}(k)$, as a function of the deformation wave vector $k$. A
negative value of ${\cal E}(k)$ signals an instability at the
corresponding wavelength, leading to phase diagrams as in
Figs.~\ref{pdsf} and \ref{pdas}. In particular, both in the
presence of salt and in salt-free conditions, (see
Figs.~\ref{pdsf} and \ref{pdas}), we find that counterion
fluctuations {\it cannot} trigger collapse of a stiff
polyelectrolyte, {\it unless} it has a microscopic persistence
length less than a critical value of
\begin{equation}
\ell_p^c=\Delta_{i} \times a \times z^4 \times \left({\ell_B \over
a}\right)^3
                \times \left(1-{a \over z \ell_B}\right)^2.
       \label{lc1}
\end{equation}
The index $i$ refers to either added-salt ($as$) or salt-free ($sf$) conditions,
with corresponding numerical constants  $\Delta_{as}$ and $\Delta_{sf}$
which are given later.
For $\ell_p < \ell_p^c$, there is a finite domain of
intermediate polyelectrolyte lengths $L$, for which a collapsed
conformation is favored.

For a low concentration of added salt, we find an effective
persistence length of
\begin{equation}
L_p=\ell_p+\frac{a \left(a / \ell_B \right)^3}{16 z^4
 \left(1-{a \over z \ell_B}\right)^2 (\kappa a)^2
\ln^2\left({1 \over \kappa a}\right)} -\frac{c_2}{2 \kappa
\ln\left(1 \over \kappa a\right)},  \label{Lpeff}
\end{equation}
where $\kappa^{-1}$ is the Debye screening length, related to the salt
(number) density $n$ via $\kappa^2=4 \pi \ell_B n$, and $c_2$ is a
numerical constant given below. The above expression, which is a {\it
  non-trivial} generalization of the Odijk-Skolnick-Fixman (OSF)
electrostatic persistence length \cite{OSF}, is a sum of repulsive (+)
and attractive (-) electrostatic contributions. It can therefore be
negative under certain conditions, which we take as an indication of a
conformational instability of the rodlike polyelectrolyte (tendency to
collapse). This occurs in the regions of the phase diagram indicated
in Fig. \ref{pdas}, and only for persistence lengths less than a
critical value given by Eq.~(\ref{lc1}) above.  A similar `softening'
contribution to the rigidity and an associated conformational
instability, can also be obtained for charged membranes \cite{LP}.
There are related instabilities caused by surface fluctuation--induced
interactions between stiff polymers on membranes\cite{RG}.

We also find that the onset of conformational instability is a
re-entrant phenomenon as a function of polyelectrolyte length for the
unscreened case, and the Debye length or salt concentration for the
screened case. This may be relevant in understanding the re-entrant
condensation of DNA that has been observed experimentally
\cite{re-entrant}, but note that alternative explanations for this
phenomenon also exist in the literature \cite{Shklovskii2}.

\section{The Model}  \label{sModel}

\begin{figure}
\centerline{\epsfxsize 7cm{\epsffile{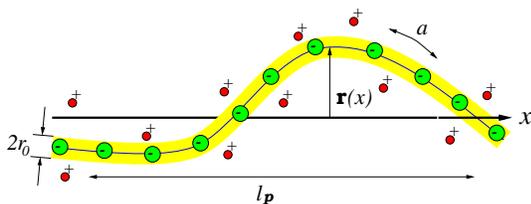}}}
\caption{Schematic of the extended polyelectrolyte parametrized by
$R(x) = (x, {\bf r}(x))$.}
% FIG.~1. Schematic of extended polyelectrolyte parametrized by $R(x) = (x, {\bf r}(x))$.
\label{schem}\end{figure}

\begin{figure}
\centerline{\epsfxsize 8cm{\epsffile{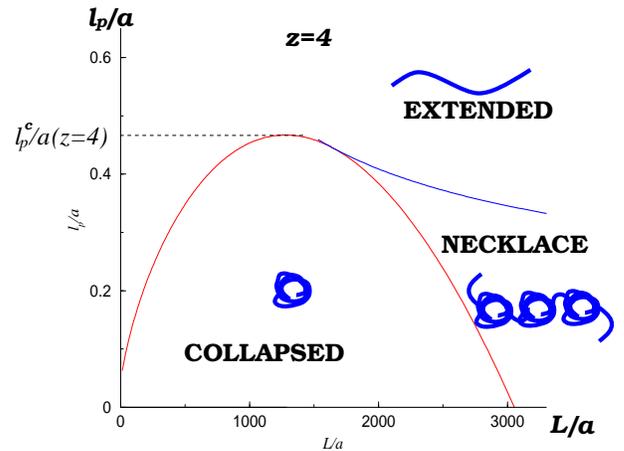}}}
\caption{Phase diagram for the salt-free case, for $z=4$. The
parameters $a=1.7 \AA$, and $k_0=10^{-3}\pi/L $ are used.}
% FIG.~2. Phase diagram for the salt-free case, for $z=4$. The parameters $a=1.7 \AA$, and $k_0=10^{-3} \pi/L$ are used.
\label{pdsf}\end{figure}

\begin{figure}
\centerline{\epsfxsize 8cm{\epsffile{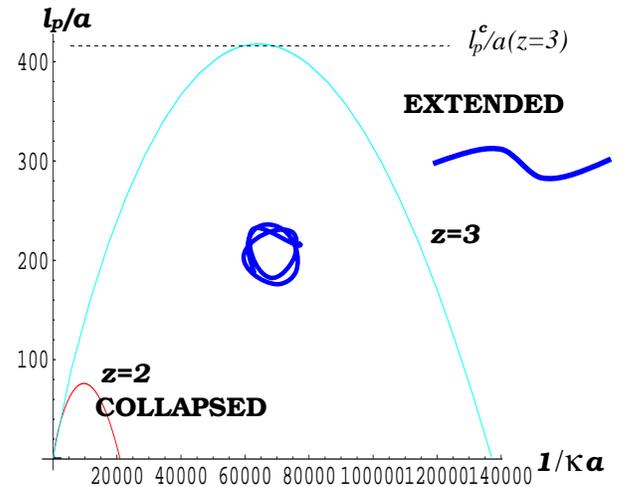}}}
\caption{Phase diagram for the added-salt case, for different
values of the counterion valence $z$.}
% FIG.~3. Phase diagram for the added-salt case, for different values of the counterion valence $z$.
\label{pdas}\end{figure}

We consider a polyelectrolyte oriented along the $x$-axis, and
parameterize is transverse deviations from a straight line by a
two component vector ${\bf r}(x)$. By assuming a single valued
function ${\bf r}(x)$, we implicitly neglect overhangs and knots,
as sketched in Fig.~\ref{schem}. The embedding of the
polyelectrolyte in space is thus described by the three component
vector $\vec{R}(x)=(x,{\bf r}(x))$. For each conformation of the
polyelectrolyte, we would like to calculate a constrained
partition function ${\cal Z}[\vec{R}]$, by integrating over all
possible configurations of the counterions in the solution. We do
this by using a phenomenological model, which generalizes previous
work by Ha and Liu~\cite{HaLiu} (and is referred to as the Ha-Liu
model henceforth), within which the restricted partition function
is written as
\begin{eqnarray}
{\cal Z}[\vec{R}]&\equiv& e^{- \beta {\cal H}^{\rm eff}[\vec{R}]}=
e^{- \beta {\cal H}^p[\vec{R}]} \times \;
\int {\cal D} q(x) \nonumber \\
&& \times \;\exp\left\{ - {1 \over 2} \int {d x \over a}
\frac{[q(x)-q_0]^2}{(\delta q)^2} \right.\nonumber \\
&&\left.-{ \ell_B \over 2} \int {d x \over a}{d x' \over a}
\frac{q(x) q(x') e^{-\kappa
|\vec{R}(x)-\vec{R}(x')|}}{\left|\vec{R}(x)-\vec{R}(x')\right|}\right\}.
\label{Z2}
\end{eqnarray}
In this equation, the energy cost for the deformations of the
polyelectrolyte is characterized by an intrinsic bending energy
described by the Hamiltonian
\begin{equation}
\beta {\cal H}^p[R]={\ell_p \over 2} \int d x (\partial_x^2
\vec{R})^2, \label{Hp}
\end{equation}
and a fluctuating electrostatic self energy. The fluctuating
linear charge distribution on the PE is characterized by a
Gaussian distribution, with the mean (corresponding to the average
renormalized charge density and read off from the exact
Fuoss-Katchalsky-Lifson solution for an infinitely long uniformly
charged rod \cite{PBrod}) given by $q_0=a/z \ell_B$, and the
variance given by $(\delta q)^2=z (1-a/z \ell_B)$ \cite{HaLiu}. It
should be noted that we have included the effect of a low density
of added salt in the above equation by adding a screening term
\cite{fisher}. The salt-free case corresponds to $\kappa = 0$.

\section{Effective elasticity of the polyelectrolyte}     \label{sEffE}

We next use the methods of Refs. \cite{KGRMP,LiK}, to calculate an
effective Hamiltonian from the above Gaussian path integral,
perturbatively in the deformation field ${\bf r}(x)$. We can
rewrite the expression in Eq.~(\ref{Z2}) above as
\begin{eqnarray}
{\cal Z}[\vec{R}]&=&e^{- \beta {\cal H}^p[\vec{R}]-{L q_0^2 \over
2 a (\delta q)^2}} \times \; \int {\cal D} q(x) \nonumber
\\
&& \times \; \exp\left\{ - {1 \over 2} \int {d x \over a}{d x'
\over a} q(x) M(x,x') q(x')\right.\nonumber \\
&&\left.\;\;\;\;\;\;\;\;\;\;\;+{q_0 \over (\delta q)^2}\int {d x
\over a} q(x) \right\}, \label{Z31}
\end{eqnarray}
where
\begin{eqnarray}
M(x,x')={a \over (\delta q)^2}\delta(x-x')+\frac{\ell_B \;
e^{-\kappa
|\vec{R}(x)-\vec{R}(x')|}}{\left|\vec{R}(x)-\vec{R}(x')\right|}.
\label{M(x,x')}
\end{eqnarray}
%Note that the geometric $\sqrt{1+(\partial_x {\bf r}(x))^2}$
%factors are necessary to ensure rotational symmetry.
Functional integration over $q(x)$ then yields
\begin{eqnarray}
\beta {\cal H}^{\rm eff}&=&\beta {\cal H}^p[\vec{R}]+{L q_0^2
\over 2 a (\delta q)^2}+{1 \over 2} \ln \det \{M(x,x')\} \nonumber
\\
&& -{1 \over 2}{q_0^2 \over (\delta q)^4} \int {d x \over a}{d x'
\over a} M^{-1}(x,x').\label{Z32}
\end{eqnarray}
Now we can write
\begin{equation}
M(x,x')=M_0(x-x')+\delta M(x,x'),   \label{deltaMx}
\end{equation}
or alternatively in Fourier space
\begin{equation}
M(k,k')=2 \pi \delta(k+k') M_0(k)+\delta M(k,k'), \label{deltaMk}
\end{equation}
in which
\begin{equation}
M_0(k)={a \over (\delta q)^2}-\ell_B \ln[(k^2+\kappa^2)
a^2],\label{M-0k}
\end{equation}
and expand Eq.~(\ref{Z32}) in powers of $\delta M$. To the leading
order, this yields
\begin{eqnarray}
\beta {\cal H}^{\rm eff}&=&\beta {\cal H}^p+\beta {\cal F}_0+{1
\over 2 }\; {\rm
tr} (M_0^{-1} \delta M) \nonumber \\
&&+{1 \over 2}{q_0^2 \over (\delta q)^4}  {\delta M(k=0,k'=0)
\over [M_0(k=0)]^2}.\label{Z33}
\end{eqnarray}
In the above equation, the trace term is an attractive
fluctuation--induced effective free energy, and the term
proportional to $q_0^2$ is a repulsive electrostatic (self-) free
energy. The constant term $\beta {\cal F}_0$ is independent of the
deformations, and thus plays no role in the effective elasticity
of the chain. It will be neglected henceforth.

Using the definition of $M$ [Eq.~(\ref{M(x,x')})], and the
expansion about the rodlike configuration, we find to the leading
order
\begin{eqnarray}
\delta M(k,-k)&=&\int {d p \over 2 \pi}|{\bf r}(p)|^2\left\{{a
\over 2 (\delta q)^2} p^2-\ell_B p^2 \ln[(k^2+\kappa^2)
a^2]\right. \nonumber \\
&&-{\ell_B \over 2} (k^2+\kappa^2) \ln[(k^2+\kappa^2)
a^2]\nonumber \\
&&+{\ell_B \over 4} [(p+k)^2+\kappa^2] \ln[(p+k)^2 a^2+(\kappa
a)^2]\nonumber \\
&&\left.+{\ell_B \over 4} [(p-k)^2+\kappa^2] \ln[(p-k)^2 a^2
+(\kappa a)^2]\right\}. \label{deltaMk2}
\end{eqnarray}
We can then use the expressions for $M_0(k)$ [Eq. (\ref{M-0k})]
and $\delta M(k,-k)$ [Eq. (\ref{deltaMk2})] and put them into
Eq.~(\ref{Z33}) to find the effective Hamiltonian. We find
\begin{equation}
\beta {\cal H}^{\rm eff}[{\bf r}(x)]={1 \over 2} \int {d k \over 2
\pi}\;{\cal E}(k)\; |{\bf r}(k)|^2 +O(r^4), \label{Fns}
\end{equation}
where
\begin{eqnarray}
{\cal E}(k)&=&\ell_p k^4 \nonumber \\
&+&{q_0^2 \ell_B \over 2 a^2}
\left\{{(k^2+\kappa^2) \ln[(k^2+\kappa^2) a^2]-\kappa^2 \ln[
(\kappa a)^2] \over \left[1+2\left(\ell_B \over a\right)(\delta
q)^2
\ln\left(1 \over \kappa a\right)\right]^2}\right\}\nonumber \\
&-&{\ell_B \over 2 a}(\delta q)^2 \int {d p \over 2 \pi}
\left\{\frac{(p^2+\kappa^2) \ln[(p^2+\kappa^2)
a^2]}{1-\left(\ell_B \over a\right)(\delta
q)^2 \ln[(p^2+\kappa^2) a^2]}\right. \nonumber \\
&&\left.-\frac{[(k+p)^2+\kappa^2] \ln[(k+p)^2 a^2 +(\kappa a)
^2]}{1-\left(\ell_B \over a \right)(\delta q)^2 \ln[(p^2+\kappa^2)
a^2]}\right\}+\tau k^2. \label{E0}
\end{eqnarray}

The above expression for the elastic kernel consists of four
different contributions:

(i) The intrinsic rigidity of the polyelectrolyte, which has a regular $k^4$
dependence corresponding to curvature elasticity.

(ii) A repulsive contribution corresponding to electrostatic
stiffening of the polyelectrolyte. This is a generalization of the wave vector
dependent rigidity first introduced by Barrat and Joanny
\cite{BJ1,BJ2}, and has a non-analytical $k$ dependence for $\kappa=0$.
(In fact, one can simply recover their formula by setting $(\delta q)^2=0$.)

(iii) A fluctuation--induced attractive term that competes with
the previous two contributions and tends to soften the polyelectrolyte.

(iv) A term corresponding to line tension renormalization with a
$k^2$ dependence. In general, one can show that such a term should
be absent since the rotational symmetry of the original
Hamiltonian for a polyelectrolyte requires ${\cal E}(k)/k^2 \to 0$ in the limit
$k \to 0$ \cite{KK}. Since the other terms in Eq.~(\ref{E0}) above
have a non-vanishing limit as $k \to 0$, one can simply tune $\tau$
in such a way that the overall line tension vanishes in that limit\cite{Tom}.

We now examine two limiting cases of the above result in more
details.

\subsection{Salt-free solution}

The salt-free case is relevant to situations in which the
screening length is much larger than the length of the polyelectrolyte, i.e.,
for $\kappa^{-1} \gg L$. Also in this limit the deformation energy is a
non-analytic function of $k$, and taking the $\kappa=0$ limit of Eq.~(\ref{E0})
requires some care.
\begin{equation}
{\cal E}(k)= \frac{\left(q_0^2 \over a^2 \right) \ell_B\; k^2
\ln\left[\left(k \over k_0 \right)^2 +1 \right]} {2
\left[1+2(\delta q)^2\left(\ell_B \over a \right)\ln\left(L \over
a \right)\right]^2} -B_1 |k|^3+\ell_p k^4, \label{E1}
\end{equation}
with $k_0 \ll \pi/L$ being a cutoff which in the absence of
screening is required for a polyelectrolyte \cite{k0}. Note that
$\partial_{k^2} {\cal E}(k)|_{k \rightarrow0}=0$ to ensure the
rotational symmetry, and that Eq.~(\ref{E1}) is not just the first
few terms in a low-$k$ expansion. The constant $B_1$ is
\begin{equation}    \label{B1eq}
B_1=\int_0^{L/a} {d x \over 4\pi} {(1+x^2)\ln\left|x^2 \over
x^2-1\right|-2 x\ln\left|x+1 \over x-1\right|+3 \over {(a/\ell_B)
\over 2 (\delta q)^2}+\ln\left({L/a \over x}\right)}.
\end{equation}
The integral $B_1$ very slowly depends on $(\delta q)^2(\ell_B/a)$
for typical values, and can be best approximated as $B_1 \simeq
c_1/\ln^2\left(L \over a\right)$ with $c_1\simeq 0.101$.

\begin{figure}
\centerline{\epsfxsize 8cm{\epsffile{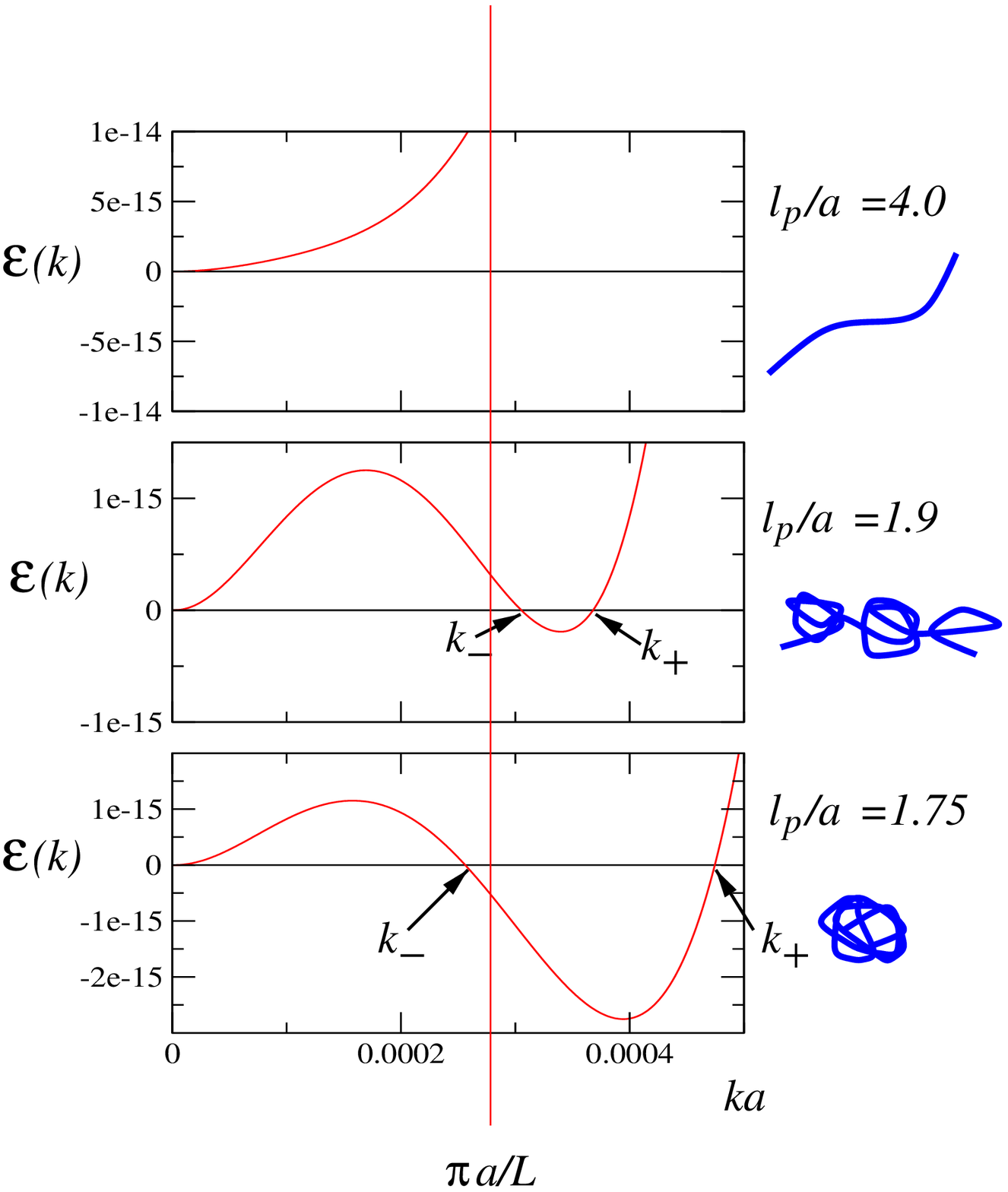}}} \caption{The
energy spectrum ${\cal E}(k)$ for the salt-free regime showing the
unstable modes.} \label{sf_spect}\end{figure}

The spectrum of the deformation modes in Eq.~(\ref{E1}) consists
of an electrostatic repulsion term (that yields the familiar
spectrum of a flexible polyelectrolyte in the limit $(\delta
q)^2=0$ \cite{KK,LiWi}). The fluctuation-induced attraction
reduces the energy cost, and could even lead to an instability at
shorter length scales\cite{k^3}. The rigidity term  $\ell_p k^4$
is the largest power of $k$ included in Eq.~(\ref{E1}) and ensures
stability at the shortest scales. A negative value of ${\cal
E}(k)$ for any $k$ indicates a linear instability; the onset of
which can be tracked by finding the point when when the minimum of
the spectrum, determined from ${d{\cal E}(k)/d k}=0$, hits the
line ${\cal E}(k)=0$. This corresponds to the criterion
$\ell_p=\ell_p^c$, where the critical persistence length
$\ell_p^c$ is given by Eq.~(\ref{lc1}) above, with
\begin{equation}    \label{Deltasf}
\Delta_{\rm sf}={c_1^2 \over \ln^2\left(L \over a\right)
\ln\left[{c_1 \over 2 \ell_p k_0 \ln^2(L/a)}\right]}.
\end{equation}
For $\ell_p < \ell_p^c$, there is domain of unstable modes for
$k_{-} < k < k_{+}$, where
\begin{equation}    \label{kpm}
k_{\pm}={c_1 \over 2 \ell_p \ln^2(L/a)} \left[1 \pm
\sqrt{1-\ell_p/\ell_p^c}\right].
\end{equation}
The spectrum of modes given by Eq.~(\ref{E1}) is plotted in
Fig.~\ref{sf_spect}. To determine the effect of the unstable modes
on the conformation of the polyelectrolyte, we should compare the
wave vectors with $\pi/L$. There are three possibilities: (i) For
$k_{+} < \pi/L$, the unstable modes cannot be accessed and the
polyelectrolyte has an {\it extended} structure. (ii) For $k_{-} <
\pi/L < k_{+}$, long wavelength modes have negative energy. Their
unstable growth is likely to lead to the collapse of the whole
chain. (iii) For $\pi/L < k_{-}$, the longest wavelength $(\sim
\pi/L)$ are stable, but there is a range of unstable wavelengths.
While linear stability cannot determine the eventual state of the
polyelectrolyte, the presence of a finite length scale could well
be the precursor of a  {\it necklace} structure.

The corresponding phase diagram is depicted in Fig.~\ref{pdsf}. A
suggested boundary between the collapsed and other phases is
obtained from $k_{\pm}=\pi/L$, as
\begin{equation}
{\ell_p \over a}=\frac{\left(L \over a\right)}{\pi \ln^2\left(L
\over a \right)}\left[c_1 -\frac{\left(L \over a \right)
\ln\left(\pi \over L k_0 \right)}{4\pi z^4 \left(\ell_B \over a
\right)^3 \left(1-{a \over z \ell_B}\right)^2}\right], \label{ph1}
\end{equation}
and has a nearly parabolic shape, leading to reentrant extended states.

\subsection{Added-salt}

The addition of salt leads to a finite screening length $\kappa^{-1}$,
and for  $\kappa^{-1} \ll L$, the spectrum ${\cal E}(k)$ given by
Eq.~(\ref{Fns}) is an analytic function of $k$ with a well defined power expansion.
The lowest order term in the expansion is proportional to $k^4$, whose
coefficient can be regarded as the `effective rigidity' of the polyelectrolyte,
and given by
\begin{equation}
L_p=\ell_p+{1 \over 4}\frac{q_0^2 \ell_B} {\left[1+2(\delta q)^2
\left(\ell_B \over a \right) \ln\left(1 \over \kappa
a\right)\right]^2 (\kappa a)^2} -{B_2\over 2 \kappa}, \label{Lp2}
\end{equation}
in which
\begin{eqnarray}    \label{B2eq}
B_2&=&\int_0^{1/\kappa a} {d x \over 4\pi} {11 x^4+2 x^2-1 \over
(x^2+1)^3 \left\{{\left(a/\ell_B\right) \over (\delta
q)^2}+\ln\left[{\left(1/\kappa a\right)^2+1 \over
x^2+1}\right]\right\}}.
\end{eqnarray}
Again, the integral $B_2$ has a very slow dependence on $(\delta
q)^2(\ell_B/a)$ for typical values, and can be best approximated
as $B_2 \simeq c_2/\ln\left(1 \over \kappa a\right)$ with
$c_2\simeq 0.288$, leading to the effective persistence length
given in Eq.~(\ref{Lpeff}). It is interesting to note that a
similar result has been predicted for stiff polyampholytes
\cite{HaThi}.

Interestingly, Eq.~(\ref{Lp2}) reproduces the OSF electrostatic
persistence length in the limit $(\delta q)^2=0$ \cite{OSF}, with
a reduced charge density $q_0$. Upon including the counterion
fluctuations, $(\delta q)^2 \neq 0$, there is a
reduction of the repulsive term, as well as the appearance of an
attractive term. As a result, the polyelectrolyte `rigidity'
(effective persistence length) can become very low and even have
negative values which we take as indicating a conformational
instability (collapse). Using this criterion ($L_p=0$), we obtain the phase
diagram shown in Fig.~4,
%Fig.~\ref{pdsa}.
which like the salt--free case, has
a nearly parabolic shape, with a maximum that yields the
critical persistence length given in Eq.~(\ref{lc1}) with
$\Delta_{\rm as}=c_2^2\simeq 0.0829$.
Once more, its shape suggests that the instability is a re-entrant effect.
Note that the critical persistence length for the case with added
salt is larger than the salt--free case.

\section{Discussion}    \label{sDisc}

The above results can be related to some experimental findings
concerning polyelectrolyte elasticity in the presence of multivalent
counterions. In a direct single DNA manipulation experiment, Baumann
{\em et al.} used a force-measuring optical tweezer to determine the
elastic properties of $\lambda$-bacteriophage DNA in the presence of
multivalent counterions \cite{Baumann}.  They observed that the
apparent persistent length of the DNA (extracted from force--extension
curves using the wormlike chain model) goes down to as low as 250-300
$\AA$, which is well below the fully saturated high salt value 450-500
$\AA$ (denoted as $\ell_{p}$ above) in the presence of multivalent
ions\cite{Baumann}.  This is in agreement with the generalised
effective persistence length given in Eq.~(\ref{Lpeff}). There may be
of course, other contributions to the effective persistence length
coming from other types of charge correlations.

Flexible polyelectrolytes have also been shown to collapse in the
presence of divalent ions~\cite{huber}. Interestingly, there is
evidence that for sodium polyacrylate chains the collapsed states
are not always compact spherical shapes at {\em lower} salt
concentrations~\cite{huber} and may have cigar-like or pearl-necklace
shapes. This is in agreement with the results we obtained in the
salt-free regime.

We now conclude with a qualitative summary of the nature of the
results, and the range of their validity. From dimensional
analysis, it is easy to show that {\it unscreened} Coulomb
interactions make a contribution of $\ell_B (k/a)^2$ to the
rigidity spectrum ${\cal E}(k)$. In the PB solution, due to charge
condensation the strength of this term is reduced by a factor of
$q_0^2=(a/z\ell_B)^2$. If the charge density on the polyelectrolyte is allowed
to fluctuate, it is further reduced, and the Coulomb rigidity goes
down by a factor of $(\delta q)^4(\ell_B/a)^2\ln^2(L/a)$, with
$(\delta q)^2=z(1-q_0)$. However, these reductions do not change
the overall sign which still prefers a rodlike structure. An
attractive (destabilizing) contribution is generated by
fluctuation-induced interactions, which are typically independent
of microscopic parameters, and hence make a contribution of $-k^3$
to ${\cal E}(k)$. In comparison to the leading Coulomb contribution,
the latter corrections become important at {\it short scales} of
order of $a^2/\ell_B$. We thus may well question the applicability
of continuum formulations to describe such a short-distance
instability. In hindsight, the phase diagrams of Fig.~2 and Fig.~3
indicate that the prefactors involved in softening of the residual
repulsion conspire to make the actual instability lengths quite
large ($\sim z^4 \ell_B^3/a^2$), and thus the continuum
formulation should hold for a large portion of these phase
diagrams \cite{z=1}.

Another potential concern is the choice of the charge variance
$(\delta q)^2$. While our approach ignores the finite size of the
counterions and allows for a large number of them to be condensed
in the vicinity of a single charged group on the polyelectrolyte,
there are other models that have restricted this number to one
\cite{Monica,Solis}. Given the relatively large radius of DNA and
actin as compared to the size of typical counterions, the
restriction of this number to one seems to be artificial, as it
leads erroneously to zero charge fluctuations at complete
condensation. Nonetheless, we believe that considering the finite
size of the counterions in a realistic way will certainly be a
worthwhile improvement of our approach. Finally, the instability
analysis performed here only provides us with information
concerning the onset of a conformational change. The final
structure of the collapsed chain is naturally beyond this linear
stability analysis\cite{Bloomfield,Gelbart}.

It is known experimentally that different counterions with the same
valence may behave differently as collapsing agents. The difference is
usually attributed to modifications in the microscopic structure of
the polyelectrolyte that take place upon binding of the
counterions\cite{Bloomfield}. It is thus plausible that the
microscopic features that distinguish between different counterions
with the same electrostatic properties can be encoded in a single
parameter, the microscopic persistence length, which normally depends
only on the local microscopic structure of the polyelectrolyte
backbone.

\acknowledgments

We have benefited from many helpful discussions with A.Yu.
Grosberg, K. Kremer, and C.R. Safinya. We would like to express
our gratitude to M. Kardar for a critical reading of the
manuscript and many useful comments. Financial support from the
Royal Society and the European Union under Marie Curie research
grant FMBICT972699 (TBL), and the Max-Planck-Gesellschaft is also
gratefully acknowledged.

%\appendix

%\end{multicols}

\end{multicols}

\begin{references}

\bibitem{Man}
G.S. Manning, J. Chem. Phys. {\bf 51}, 954 (1969).

\bibitem{Oos}
F. Oosawa, Biopolymers {\bf 6}, 134 (1968); F. Oosawa, {\it Polyelectrolytes}
(Marcel Dekker, New York, 1971).

\bibitem{BJ1}
J.L. Barrat and J.F. Joanny, Adv. Chem. Phys. {\bf 94}, 1 (1996).

\bibitem{LeBret}
M. Le Bret, J. Chem. Phys. {\bf 76}, 6243 (1982); M. Fixman, J.
Chem. Phys. {\bf 76}, 6346 (1982).

\bibitem{Delsanti}
M. Delsanti, J.P. Dalbiez, O. Spalla, L. Belloni, and M. Drifford,
ACS Symp. Ser. {\bf 548}, 381 (1994).

\bibitem{huber}
R. Schweins and K. Huber, Eur. Phys. J. {\bf E 5}, 117
(2001); Y. Ikeda, M. Beer, M. Schmidt and K. Huber, Macromolecules
{\bf 31}, 728 (1998); K. Huber, J. Phys. Chem. {\bf 97}, 9825
(1993). There are also similar observations in: B.L. Smith, D.R.
Gallie, H. Le, and P.K. Hansma, Journal of Structural Biology {\bf
119}, 109 (1997).

\bibitem{SK}
M.J. Stevens and K. Kremer, Phys. Rev. Lett. {\bf 71}, 2228
(1993); J. Chem. Phys. {\bf 103}, 1669 (1995).

\bibitem{Monica}
P. Gonzalez-Mozuelos and M. Olvera de la Cruz, J. Chem. Phys. {\bf
103}, 3145 (1995); M. Olvera de la Cruz, L. Belloni, M. Delsanti,
J.P. Dalbiez, O. Spalla, M. Drifford, J. Chem. Phys. {\bf 103},
5781 (1995).

\bibitem{BKK}
N.V. Brilliantov, D.V. Kuznetsov, and R. Klein, Phys. Rev. Lett.
{\bf 81}, 1433 (1998).

\bibitem{SchP}
H. Schiessel and P. Pincus, Macromolecules {\bf 31}, 7953 (1998).

\bibitem{Bloomfield}
V.A. Bloomfield, Biopolymers {\bf 31}, 1471 (1991); V.A. Bloomfield, Curr. Opin.
Struct. Biol. {\bf 6}, 334 (1996).

\bibitem{Bundle}
J.X. Tang, S. Wong, P. Tran, and P.A. Janmey, Ber. Bunsen-Ges.
Phys. Chem. {\bf 100}, 1 (1996); J.X. Tang, T. Ito, T. Tao, P.
Traub, and P.A. Janmey, Biochemistry {\bf 36}, 12600 (1997).

\bibitem{lyubart}
A.P. Lyubartsev, J.X. Tang, P.A. Janmey, and L. Nordenskiold,
Phys. Rev. Lett. {\bf 81}, 5465 (1998).

\bibitem{stevens}
M.J. Stevens, Phys. Rev. Lett. {\bf 82}, 101 (1998).


\bibitem{Man2}
G.S. Manning, Biopolymers {\bf 19}, 37 (1980); G.S. Manning, Cell
Biophys. {\bf 7}, 57 (1985).

\bibitem{Rouzina}
I. Rouzina, and V. A. Bloomfield, Biophys. J. {\bf 74}, 3152
(1998).

\bibitem{Nguyen}
T. T. Nguyen, I. Rouzina, and B. I. Shklovskii, Phys. Rev. E
{\bf 60}, 7032 (1999).

\bibitem{GKL}
R. Golestanian, M. Kardar, and T.B. Liverpool, Phys. Rev. Lett.
{\bf 82}, 4456 (1999).

\bibitem{Podgornik}
P.L. Hansen, R. Podgornik, D. Svensek, and V. A. Parsegian, Phys.
Rev. E {\bf 60}, 1956 (1999).

\bibitem{Andelman}
G. Ariel and D. Andelman, preprint cond-mat/0112337.

\bibitem{Rouzina2}
I. Rouzina and V.A. Bloomfield, J. Phys. Chem. {\bf 100}, 9977
(1996).

\bibitem{Gron}
N. Gronbech-Jensen, R.J. Mashl, R.F. Bruinsma, and W.M. Gelbart,
Phys. Rev. Lett. {\bf 78}, 2477 (1997)


\bibitem{Shklovskii}
B.I. Shklovskii, Phys. Rev. Lett. {\bf 82}, 3268 (1999).

\bibitem{Levin}
J.J. Arenzon, J.F. Stlick, and Y. Levin, Eur. Phys. J. B {\bf 12},
79 (1999).

\bibitem{Solis}
F.J. Solis and M. Olvera de la Cruz, Phys. Rev. E {\bf 60}, 4496
(1999).

\bibitem{Leikin}
A.A. Kornyshev, and S. Leikin, J. Chem. Phys. {\bf 107}, 3656
(1997); Biophys. J. {\bf 75}, 2513 (1998); Phys. Rev. Lett. {\bf
82}, 4138 (1999); Phys. Rev. Lett. {\bf 84}, 2537 (2000).

\bibitem{Pod1}
R. Podgornik and V.A. Parsegian, Phys. Rev. Lett. {\bf 80}, 1560
(1998).

\bibitem{HaLiu}
B.-Y. Ha and A.J. Liu, Phys. Rev. Lett. {\bf 79}, 1289 (1997);
Phys. Rev. Lett. {\bf 81}, 1011 (1998); Physica A {\bf 259}, 235
(1998); Phys. Rev. E {\bf 58}, 6281 (1998); Europhys. Lett. {\bf
46}, 624 (1999); Phys. Rev. E {\bf 60}, 803 (1999).

\bibitem{KGRMP}
M. Kardar and R. Golestanian, Rev. Mod. Phys. {\bf 71}, 1233
(1999).

\bibitem{overcharge}
B.I. Shklovskii, Phys. Rev. E {\bf 60}, 5802 (1999).

\bibitem{messina}
R. Messina, C. Holm, and K. Kremer, Phys. Rev. Lett. {\bf 85}, 872
(2000); Europhys. Lett. {\bf 51}, 461 (2000).

\bibitem{ray}
J. Ray and G.S. Manning, Langmuir {\bf 10}, 2450 (1994);
Macromolecules {\bf 30}, 5739 (1997); Macromolecules {\bf 33},
2901 (2000);

\bibitem{Roland}
A.G. Moreira and R.R. Netz, Phys. Rev. Lett. {\bf 87}, 078301
(2001); R.R. Netz, Eur. Phys. J. E {\bf 5}, 557 (2001); R.R. Netz,
in {\it Electrostatic Effects in Soft Matter and Biophysics}, C.
Holm et al. (eds.), pp. 367-408 (Kluwer, Amsterdam, 2001).

\bibitem{Andy}
A.W.C. Lau, P. Pincus, D. Levine, and H.A. Fertig, Phys. Rev. E
{\bf 63}, 051604 (2001).

%\bibitem{Gerard}
%Gerard Wong, private communication.

\bibitem{LiK}
H. Li and M. Kardar, Phys. Rev. Lett. {\bf 67}, 3275 (1991); Phys.
Rev. A {\bf 46}, 6490 (1992); R. Golestanian and M. Kardar, Phys.
Rev. Lett. {\bf 78}, 3421 (1997); Phys. Rev. A {\bf 58}, 1713
(1998).

\bibitem{OSF}
T. Odijk,, J. Polym. Sci. {\bf 15}, 477 (1977);
J. Skolnick and M. Fixman, Macromolecules {\bf 10}, 944 (1977).

\bibitem{LP}
A.W.C. Lau and P. Pincus, Phys. Rev. Lett. {\bf 81}, 1338 (1998);
B.-Y. Ha, Phys. Rev. E {\bf 64}, 031507 (2001); Phys. Rev. E {\bf
64}, 051902 (2001); R.R. Netz, Phys. Rev. E {\bf 64}, 051401
(2001).

\bibitem{RG}
R. Golestanian, Europhys. Lett. {\bf 36}, 557 (1996).

\bibitem{re-entrant}
J. Pelta, D. Durand, J. Doucet, and F. Livolant, Biophysical
Journal {\bf 71}, 48 (1996); J. Pelta, F. Livolant, and J.L.
Sikorav, Journal of Biological Chemistry {\bf 271}, 5656 (1996);
E. Raspaud, M. Olvera de la Cruz, J.L. Sikarov, and F. Livolant,
Biophysical Journal {\bf 74}, 381 (1998); M. Saminathan, T.
Antony, A. Shirahata, L. Sigal, T. Thomas, and T.J. Thomas,
Biochemistry {\bf 38}, 3821 (1999).

\bibitem{Shklovskii2}
T.T. Nguyen, I. Rouzina, and B.I. Shklovskii, e-print
cond-mat/9908428.

\bibitem{PBrod}
R.M. Fuoss, A. Katchalsky, and S. Lifson, Proc. Natl. Acad. Sci.
USA {\bf 37}, 579 (1951).


\bibitem{fisher}
The density of the salt $n$ should be low enough not to smear out
the condensation. The requirement of the mass contrast to be
substantial, yields $n \ll z (1-a/z \ell_B)/(a S)$, where $S$ is
the cross sectional area of the polyelectrolyte.

\bibitem{BJ2}
J.L. Barrat and J.F. Joanny, Europhys. Lett. {\bf 24}, 333 (1993).

\bibitem{KK}
Y. Kantor and M. Kardar, Europhys. Lett. {\bf 9}, 53 (1989).

\bibitem{Tom}
In order to actually calculate $\tau$ from the path integral
formulation correctly, and prove that it does cancel the other
contributions to the line tension as rotational symmetry requires,
one should construct a more delicate field theory with invariant
measures, etc., which was not the aim of the present work. For a
related discussion in the case of fluid membranes see: W. Cai,
T.C. Lubensky, P. Nelson, and T. Powers, J. Phys. II France {\bf
4}, 931 (1994).

\bibitem{k0} For wavelength such that $ k \gg k_0$ we can approximate
$ \ln[(k/k_0)^2+1]$ by $2 \ln (k/k_0)$.
 Although, strictly
speaking, one should finally take the limit $k_0 \to 0$, it is
sufficient for practical purposes that $k_0$ is small enough such
that $\sin(k_0 L) \simeq k_0 L$. With this in mind, we have
selected  $k_0=10^{-3} \pi/L$ for the phase diagram of Fig.~2.

\bibitem{LiWi}
H. Li and T.A. Witten, Macromolecules {\bf 28}, 5921 (1995).

\bibitem{k^3}
The repulsive part of the spectrum comes from a pair potential of
the form $(a/z \ell_B)^2/z^2 \ell_B r$, while the attractive
fluctuation-induced part can be obtained from a $-1/r^2$ pair
potential. This leads to a short length scale instability for $r<
z^4 (\ell_B/a)^2 \ell_B$.

\bibitem{necklace}
A necklace structure for randomly charged polymers was introduced
in Y. Kantor and M. Kardar,  Europhys. Lett. {\bf 27}, 643 (1994);
Y. Kantor and M. Kardar, Phys. Rev. {\bf E51}, 1299 (1995). Its
applicability to uniformly charged polyelectrolytes is discussed in A.V.
Dobrynin, S.P. Obukhov, and M. Rubinstein, Macromelecules {\bf
29}, 2974 (1996).

\bibitem{HaThi}
B.-Y. Ha and D. Thirumalai, J. Phys. II (France), {\bf 7}, 887
(1997).

\bibitem{Baumann}
C.G. Baumann, S.B. Smith, V.A. Bloomfield, and C. Bustamante,
Proc. Natl. Acad. Sci. USA {\bf 94}, 6185 (1997).

\bibitem{z=1}
 From these arguments, it is clear that counterions with higher
valence $z$ are much more effective in collapsing the polyelectrolytes.
Nonetheless, Eq.(\ref{lc1}) does not rule out the possibility of
collapse for monovalent counterions, provided that the polyelectrolyte has a
low enough microscopic persistence length.

%\bibitem{Coalson}
%R.D. Coalson and A. Duncan, J. Chem. Phys. {\bf 97}, 5653 (1992).

\bibitem{Gelbart}
S.Y. Park, D. Harries, and W. M. Gelbart, Biophys. J. {\bf 75},
714 (1998).

%\bibitem{mono}
%N.A. Kasyanenko, A.V. Zanina, O.V. Nazarova, and E.F. Panarin, preprint (1998).


\end{references}
\end{document}